\documentclass[letterpaper]{aa}
\usepackage{natbib}

\newcommand\bb[1] {   \mbox{\boldmath{$#1$}}  }
\newcommand\del{\bb{\nabla}}
\newcommand\bcdot{\bb{\cdot}}

\def\dd{\partial}

\def\beq{ \begin{equation} }
\def\eeq{ \end{equation} }
\def\spose#1{\hbox to 0pt{#1\hss}} % from Scott Tremaine
\def\ltsim{\mathrel{\spose{\lower.5ex\hbox{$\mathchar"218$}}
\raise.4ex\hbox{$\mathchar"13C$}}}

\begin{document}

\title{Nonlinear Scale Invariance in Local Disk Flows}

\author{Steven A. Balbus\inst{1}\inst{,2} }

\institute{Dept.~of Astronomy, University of Virginia, Charlottesville,
VA 22903, USA, \and Ecole Normale Sup\'erieure, Laboratoire de Radioastronomie,
24, rue Lhomond, 75231, Paris, France}

\offprints{Steven A. Balbus,
\tt{sb@virginia.edu}}               

%\email {sb@virginia.edu}

\authorrunning{Balbus}
%\titlerunning{Local stability}

\date{Received ; accepted }

\abstract{
An exact nonlinear scaling transformation is presented for the local
three-dimensional dynamical equations of motion for differentially
rotating disks.  The result is relevant to arguments that have been put
forth claiming that numerical simulations lack the necessary numerical
resolution to resolve nonlinear instabilities that are supposedly present.
We show here that any time dependent velocity field satisfying the local
equations of motion and existing on small length scales, has an exact
rescaled counterpart that exists on arbitrarily larger scales as well.
Large scale flows serve as a microscope to view small scale behavior.  The
absence of any large scale instabilities in local numerical simulations
of Keplerian disks suggests that the equations in this form have no
instabilities at any scale, and that finite Reynolds number suppression
is not the reason for the exhibited stable behavior.  While this argument
does not rule out the possibility of global hydrodynamical instability,
it does imply that differential rotation {\em per se} is not unstable in a
manner analogous to shear layers or high Reynolds number Poiseuille flow.
Analogies between the stability behavior of
accretion disks and these flows are specious.
 }

\maketitle

\section{Introduction}
Disks with Keplerian rotation profiles are linearly stable by the
Rayleigh criterion of outwardly increasing specific angular momentum,
but are extremely sensitive to the presence of magnetic fields.
A weakly magnetized disk is linearly unstable if its angular
velocity decreases outward, a condition met by Keplerian and almost
all other astrophysical rotation profiles (Balbus \& Hawley 1991).
The underlying physics behind this magnetorotational instability (MRI)
is well-understood, and the breakdown of the flow into fully developed
turbulence has been convincingly demonstrated in a large series of
numerical simulations (Balbus 2003 for a review).

Not all astrophysical disks need have the requisite minimum ionization
level to sustain magnetic coupling, however.  Protostellar disks, for
example, may have an extended ``dead zone'' near the midplane on radial
scales from $\sim 0.1$ to $\sim 10$ AU (Gammie 1996; Fromang, Terquem,
\& Balbus 2001).  This, along with other similar cases (e.g. CV disks,
cf. Gammie \& Menou 1998), has led to speculation that there are also 
hydrodynamical mechanisms by which Keplerian flow is destabilized
(Gammie 1996).  

Before the advent of the MRI, such reasoning was orthodox.  The pioneering
work of Shakura \& Sunyaev (1973), for example, invoked nonlinear, high
Reynolds number shear instabilities as a likely destabilizing mechanism
that would lead to turbulence (see also Crawford \& Kraft 1956).  Since,
for nonaxisymmetric disturbances, there is still no proof either of linear
or nonlinear stability, this mechanism continues to attract adherents
(Dubrulle 1993, Richard \& Zahn 1999, Richard 2003).

Theoretical analysis may have hit an impasse, but the intervening
years have in fact seen a stunning rise in the capabilities of numerical
simulation.  These have shown no indication of local nonlinear rotational
instabilities (Hawley, Balbus, \& Winters 1999), in Keplerian disks.
They do, however, reveal nonlinear shear instabilities when Corilois
forces are absent, or when the disk is marginally stable (constant
specific angular momentum).  Indeed, even linear instability is possible
in some Rayleigh-stable disks, provided that global physics is introduced
(Goldreich, Goodman, \& Narayan 1986; Blaes 1987), a result that has
been numerically confirmed (Hawley 1991).

The numerical stability findings have been criticized on the grounds
that the effective Reynolds number of the codes is too low, and that this
damps the nonlinear instabilities: the latter require yet-to-be resolved
spatial scales in order to reveal themselves (Richard \& Zahn 1999).
In this paper, we show that the local disk equations possess a scale
invariance that implies any solution to the governing equations must be
present on all scales.  In other words, for every small scale velocity
flow, there is an exact large scale counterpart with the same long term
stability behavior.  The absence of instability at large scale therefore
implies the absence at small scales as well.  Conversely, any true small
scale instabilities (those present in a shear layer, for example), must
also have large scale counterparts, and therefore instability should be
found even at crude numerical resolutions.  This is indeed the case.
Our findings suggest that if nonlinear hydrodynamical instabilities were
present in Keplerian disks, such unstable disturbances must involve
dynamics beyond the local approximation, and are not an inevitable
nonlinear outcome of differential rotation.

\section{The Local Approximation}

In cylindrical coordinates $(R, \phi, z)$,
the fundamental equations of motion for a flow in which viscous
effects are negligible are mass conservation
\beq\label{fun0}
{\dd\rho\over \dd t} + \del\bcdot(\rho\bb{v})= 0,
\eeq
and the dynamical equations,
\beq\label{fun1}
{\dd  v_R\over \dd t} + \bb{v}\bcdot\del v_R - {v_\phi^2\over R} = 
- {1\over\rho}{\dd P\over \dd R} - {\dd\Phi\over \dd R}
\eeq
\beq\label{fun2}
{\dd  v_\phi\over \dd t} + \bb{v}\bcdot\del v_\phi + {v_\phi v_R\over R} =
- {1\over\rho R}{\dd P\over \dd \phi}
\eeq
\beq\label{fun3}
{\dd  v_z\over \dd t} + \bb{v}\bcdot\del v_z 
= - {1\over\rho }{\dd P\over \dd z} -{\dd\Phi\over z}
\eeq
Our notation is standard: $\bb{v}$ is the velocity field,
$\rho$ the mass density, $P$ the gas pressure, and $\Phi$
is the Newtonian point mass potential for central mass $M$:
\beq
\Phi =  - {GM\over \left(R^2 + z^2\right)^{1/2}}.
\eeq
$G$ is the gravitational constant.  

The local limit consists of the following series of approximations.
First, we assume that $R$ is large and $z \ll R$, so that
\beq\label{phis}
\Phi \simeq  - {GM\over R} \left(1 - {z^2\over 2R^2}\right)
\eeq
and
\beq
 {\dd\Phi\over \dd R}  \simeq{GM\over R^2}, \quad
{\dd\Phi\over \dd z} \simeq {GM z\over R^3}
\eeq
Choose a fiducial value of $R$, say $R_0$.  Denote the angular
velocity as $\Omega(R)$ (we assume a dependence only upon $R$), and let
$\Omega_0 = \Omega(R_0)$.  We next erect local Cartesian coordinates
\beq
x = R - R_0, \ y = R_0(\phi -\Omega_0 t)
\eeq
which corotate with the disk at $R=R_0$.  Let
\beq\label{w}
\bb{w} \equiv \bb{v} - R\Omega_0 t \bb{e_\phi}
\eeq
be the velocity relative to uniform rotation
at $\Omega= \Omega_0$.  In the local approximation,
the magnitude of $\bb{w}$ is assumed to be small
compared with $R\Omega_0$.  

The undisturbed angular velocity is Keplerian,
\beq\label{kep}
\Omega^2 = {GM\over R^3}
\eeq
Substitution of equations (\ref{phis}-\ref{kep})
into equations (\ref{fun1}--\ref{fun2}) and retaining leading
order, yields
the so-called {\em local} or {\em Hill}
equations (e.g., Balbus \& Hawley 1998):
\beq\label{hill0}
{\dd\rho\over \dd t} + \del\bcdot(\rho\bb{w})= 0,
\eeq
\beq\label{hill1}
\left( {\dd \over \dd t} + \bb{w\cdot}\del\right){w_R} 
- 2\Omega w_\phi = - x{d\Omega^2\over d \ln R} - {1\over\rho}
{\dd P\over \dd x}
\eeq
\beq\label{hill2}
\left( {\dd \over \dd t} + \bb{w\cdot}\del\right){w_\phi} 
+ 2\Omega w_R = - {1\over\rho} {\dd P\over \dd  y}
\eeq
\beq\label{hill3}
\left( {\dd \over \dd t} + \bb{w\cdot}\del\right){w_z} 
= - z\Omega^2 - {1\over\rho } {\dd P\over \dd z}
\eeq
The ``0'' subscript has been dropped in the $2\Omega$ terms
in equations (\ref{hill1}) and (\ref{hill2}), and in the
derivative term on the right of equation (\ref{hill1}).
The time derivative is taken in the corotating frame,
viz.:
\beq
{\dd\ \over \dd t} \mbox{(inertial)} = 
{\dd\ \over \dd t} \mbox{(corotating)} + \Omega_0{\dd \over \dd\phi}
\eeq
Equations (\ref{hill0}--\ref{hill3}) are well known, and have been used
extensively in both numerical and analytical studies.  
The fundamental approach dates from nineteenth century treatments
of the Earth-moon system (Hill 1878).  

\section {Scale symmetry in the Hill Equations}
The local equations of motion incorporate an important
symmetry in their structure.  Let
\beq
\bb{w}(\bb{r}, t), \quad \rho(\bb{r}, t), \quad P(\bb{r}, t),
\eeq
where $\bb{r}=(x, y, z)$, be an exact solution to the 
Hill equations (\ref{hill0}--\ref{hill3}). Then, if
$\alpha$ is an arbitrary constant,
\beq
(1/\alpha)\bb{w}(\alpha\bb{r}, t), \quad \rho (\alpha\bb{r}, t),
\quad (1/\alpha^2) P(\alpha\bb{r}, t)
\eeq
is also an exact solution to the same equations.  The proof is
a simple matter of direct substitution.

An equivalent formulation of the scaling symmetry is
\beq
\bb{w}(\bb{r}/\epsilon, t)\leftrightarrow (1/\epsilon)
\bb{w}(\bb{r}, t)
\eeq
\beq
\rho (\bb{r}/\epsilon, t)\leftrightarrow 
\rho (\bb{r}, t)
\eeq
\beq
P (\bb{r}/\epsilon, t)\leftrightarrow (1/\epsilon^2) 
P (\bb{r}, t).
\eeq
In this form, with $\epsilon \ll 1$, we see that any solution of
the Hill equations that involves very small length scales has a
rescaled counterpart solution with exactly the same time dependence.
In particular, any solution corresponding to a breakdown into turbulence
must be present on both large and small scales.

The implications of this scaling symmetry are of particular importance
for understanding and testing the possible existence of local nonlinear
instabilities in Keplerian disks.  The key point is that any such
instability would have to exist not just at small scales, but at all
scales.  Finite difference numerical codes would find such instabilities,
if they existed.  Indeed, a constant specific angular momentum profile is
nonlinearly unstable, and is found to be so even at resolutions as crude
as $32^3$.  By way of contrast, local Keplerian  profiles show no evidence
of nonlinear instability at resolutions up to $256^3$, instead converge
to the same stable solution in codes with completely different numerical
diffusion properties (Hawley, Balbus, \& Winters 1999).  The argument that
small scale flow structure is somehow being repressed is simply untenable.

To see how the Reynolds number changes with scale,
assume that a flow is characterized by an effective
kinematic viscosity $\nu$.  The scaling argument we have just given applies
to inviscid equations, so we should not expect it to hold in the presence
of viscosity.  The Reynolds number associated with the small scale solution is
\beq
Re_{s} = {w \times \epsilon l \over \nu}
\eeq
The Reynolds number associated with the large scale solution is
\beq
Re_{l} = {w\times l\over \epsilon\nu }
\eeq
where $w$ here means $w(l, t)$, the value of the velocity
function evaluated at a fiducial value length $l$ and time $t$.
$Re_l =Re_s/\epsilon^2 \ll Re_{s}$ because at larger scales both the
velocity and the length scales increase by a factor of $1/\epsilon$.
In a numerical simulation, strict scaling invariance is not obeyed.
Instead, the large scale solutions approach the inviscid limit, while
their sufficiently small scale counterparts are damped.  But by behaving
nearly inviscidly, the large scale solutions capture the behavior of
the Hill system at all scales.

\section {What this Result Does Not Show}

Obviously, scale invariance does not constitute a proof of nonlinear
stability in any Keplerian flow.  There are several points we have
not covered.

First, the local approximation ignores boundary conditions.  In laboratory
flows, the fluid is always bounded by hard walls, and boundary
layers form.  A recent laboratory confirmation of the MRI also finds
finite amplitude velocity fluctuations in a magnetically stable flow,
for example.  But the source of such disturbances are boundary layers
(Sisan et al. 2004).

The Hill equations emerge in the limit $R\rightarrow\infty$, and therefore
curvature terms drop out of the analysis.  Instabilities that depend,
for example, upon inflection points or vorticity maxima in the background
rotation profile would not appear in this limit.  Nothing precludes
them from forming in the $w$ velocity profile, however, and if such
instabilities were present they should manifest on large scales as well
as small.  In any case, the criticism of the numerical simulations is
that extremely small structure is being lost, and that high Reynolds
number differential rotation is supposedly intrinsically unstable.
It is very difficult to see how large scale curvature could play an
essential destabilizing role here.  In these equations, the curvature terms
are nonsingular perturbations.  Planar Couette and Poiseuille
flows break down into turbulence without assistance from geometrical
curvature.

Our Hill analysis together with numerical simulations would also
suggest that a non-Keplerian disk with, say, $\Omega \propto R^{-1.8}$
is nonlinearly stable.  But an annulus supporting such a profile is
in fact {\em linearly} unstable (Goldreich, Goodman, \& Narayan 1986),
transporting angular momentum outward even in its linear phase.  The point
is that the annulus supports edge modes that become unstable, and these
global modes do not exist in the local approximation.  The existence of
a similar instabilities in disks found in nature cannot be ruled out,
though to date none afflicting Keplerian disks have been found.

The disk thermal structure could also be unstable, at least in principle.
Nothing presented in this work bears on these types of instabilities.

Finally, there are technical loopholes to the argument presented in
this paper.  What if the unstable solution required not just some small
scales to be resolved, but very disparate scales?  Why this should be
so is far from clear, but this possibility cannot be ruled out {\em a
priori.}  Indeed, one could imagine that a fractal structure is required
down to infinitesimal scales.  Rescaling would not bring such a solution
to larger characteristic length scales, by definition.  This solution
is obviously not characterized by a critical Reynolds number, above
which it is necessary to be seen.  The critical Reynolds number would be
infinity!  This is not the argument made by proponents of nonlinear high
Reynolds number instability.  Such a solution may remain a mathematical
possibility, but not one that can be realized in nature.

\section {Conclusion}

The local dynamics of Keplerian or other astrophysical disk profiles
can be captured by a an established formalism known as the local, or
Hill, approximation.  The resulting system of equations has an exact
scale invariance, so that any flow characterized by very small scales
has an exact large scale counterpart with same stability properties.
This feature of the Hill equations implies that finite difference codes
at available resolutions are sufficient to explore the possibility of
{\em local} nonlinear shear instabilities in astrophysical disks.
If simulations accurately describe the large scale behavior of
the Hill system, there is nothing more to uncover at small scales;
it is simply renormalized large scale behavior.  The absence of any
observed instabilities in Keplerian numerical studies, coupled with
the ready manifestation of such instabilities in local shear layers and
constant specific angular momentum systems, suggests that any putative
nonmagnetic disk instability would have to incorporate physics beyond
simple differential rotation.

\acknowledgements I thank C. Gammie, J. Hawley, K. Menou, and C. Terquem
for useful comments.  This work is supported by NASA grants NAG5-13288
and NNG04GK77G.

{}

\begin{thebibliography}{}

\bibitem[Balbus (2003)]{b03} Balbus, S. A. 2003, \araa, 41, 555

\bibitem[Balbus \& Hawley (1991)]{bh91}
Balbus, S. A. \& Hawley, J. F. 1991, ApJ, 376, 214

\bibitem[Balbus \& Hawley (1998)]{bh98}
Balbus, S. A. \& Hawley, J. F. 1998. Rev. Mod. Phys., 70, 1.

\bibitem[Blaes (1987)]{b87} Blaes, O. M. 1987, MNRAS, 227, 975

\bibitem[Crawford \& Kraft (1956)]{ck56} Crawford, J. A., \& Kraft, R.
P. 1956, ApJ, 123, 44

\bibitem[Dubrulle (1993)]{d93} Dubrulle, B. 1993, Icarus, 106, 59

\bibitem[Fromang, Terquem, \& Balbus (2001)]{ftb01}
Fromang, S., Terquem, C., \& Balbus, C. 2001, MNRAS, 329, 18

\bibitem[Gammie (1996)]{g96} Gammie, C. F. 1996, ApJ, 457, 355

\bibitem[Gammie \& Menou (1998)]{gm98} Gammie, C. F., \& Menou, K. 1998,
ApJ, 492, L75

\bibitem[Goldreich, Goodman, \& Narayan (1986)]{ggn86} Goldreich, P.,
Goodman, J., \& Narayan, R. 1986, MNRAS, 221, 339

\bibitem[Hawley (1991)]{h91} Hawley, J. F. 1991, ApJ, 381, 496

\bibitem[Hawley, Balbus, \& Winters (1999)]{hbw99} Hawley, J. F.,
Balbus, S. A., \& Winters, W. F. 1999, ApJ, 518, 394

\bibitem[Hill (1878)]{hill78} Hill, G. W. 1878, Am. J. Math., 1, 5

\bibitem[Richard (2003)]{r03} Richard, D. 2003, \aa, 408, 409

\bibitem[Richard \& Zahn (1999)]{rz99} Richard, D., \& Zahn, J.-P.
1999, A\&A, 347, 734

\bibitem[Shakura \& Sunyaev (1973)]{ss73} Shakura, N. I., \& Sunyaev, R.
A. 1973, A\&A, 24, 337

\bibitem[Sisan et al. (2004)]{setal04} Sisan, D. R., Mujica, N.,
Tillotson, W. A., Huang, Y.-M., Dorland, W., Hassam, A., Antonsen, T.
M., \& Lathrop, D. P. 2004, Phys. Rev., in press (physics/0401125). 

\end{thebibliography}
\end{document}